\documentclass[conference]{IEEEtran}
\IEEEoverridecommandlockouts
\usepackage{cite}
\usepackage{amsmath,amssymb,amsfonts}
\usepackage{algorithmic}
\usepackage{graphicx}
\usepackage{textcomp}
\usepackage{xcolor}
\usepackage{enumitem}
\def\BibTeX{{\rm B\kern-.05em{\sc i\kern-.025em b}\kern-.08em
    T\kern-.1667em\lower.7ex\hbox{E}\kern-.125emX}}
\begin{document}

\title{Probing Ring Resonator Sensor Based on Vernier Effect\\

}

\author{\IEEEauthorblockN{1\textsuperscript{st} Wenwen Zhang*}
\IEEEauthorblockA{\textit{Department of Electrical and Computer Engineering} \\
\textit{University of British Columbia}\\
Vancouver, Canada \\
wenwenzhang@ece.ubc.ca}
\and
\IEEEauthorblockN{2\textsuperscript{nd} Hao Zhang}
\IEEEauthorblockA{\textit{Department of Electrical and Computer Engineering} \\
\textit{University of British Columbia}\\
Vancouver, Canada \\
layallan@student.ubc.ca}
\and
}

\maketitle

\begin{abstract}
The Vernier effect has seen extensive application in optical structures, serving to augment the free spectral range (FSR). A substantial FSR is vital in a myriad of applications including multiplexers, enabling a broad, clear band comparable to the C-band to accommodate a maximum number of channels. Nevertheless, a large FSR often conflicts with bending loss, as it necessitates a smaller resonator radius, thus increase the insertion loss in the bending portion. To facilitate FSR expansion without amplifying bending loss, we employed cascaded and parallel racetrack resonators and ring resonators of varying radius that demonstrate the Vernier effect. In this study, we designed, fabricated, and tested multiple types of racetrack resonators to validate the Vernier effect and its FSR extension capabilities. Our investigations substantiate that the Vernier effect, based on cascaded and series-coupled micro-ring resonator (MRR) sensors, can efficiently mitigate intra-channel cross-talk at higher data rates. This is achieved by providing larger input-to-through suppression, thus paving the way for future applications.
\end{abstract}


\section{Introduction}
With the fast development of communication technology, the amount of information to be transmitted increased rapidly. Photonic devices play an important role in computing systems for wider bandwidth and more cost-efficient interconnects. Traditional interconnects are typically made of metallic materials, which introduces high energy consumption and large latency ~\cite{ zhang2009silicon}. Silicon-on-insulator (SOI) interconnection devices lead to a new field of communication~\cite{ xia2007ultra}. But it still demands larger free spectral range (FSR). To get extended FSR, flat passband response, or increased sensor sensitivity, various cascaded and parallel structures are designed to explore excellent spectra characteristics. The cascaded structure is not limited to Mach–Zehnder interferometers ~\cite{la2013ultra} or Fabry--Perot cavities and micro-ring resonators ~\cite{ song2012intensity}. By applying the vernier effect to optical waveguide sensors, the sensitivity of these sensors can be further improved.

Vernier effect based on micro-ring resonators (MRRs) has been validated effectively in increasing device performance in communication equipment~\cite{griffel2000vernier}. Adopting the vernier effect in ring/racetrack structures could extend the FSR which benefits dense wavelength-division multiplexing (DWDM) applications and increase sensor sensitivity ~\cite{chaichuay2009serially}. Sensors with sensitivity-enhanced refractive index are one of the outstanding applications of micro-fibers ~\cite{lu2019sensitivity}. This fiber sensor achieves sensitivity enhancement of 3301.76 nm/RIU and is implemented by connecting two microfiber knot resonators in series, which exhibits the vernier effect. Refractive index sensor based on silicon-on-insulator with vernier effect and enhanced sensitivity is also realized in ~\cite{claes2010experimental}. These sensors perform a significant role in measurements on monitoring the environment and medical analysis, food detection, etc. Silicon-on-insulator racetrack resonators with extended FSR are also realized by cascaded rings with exhibit vernier effect ~\cite{boeck2010series}. Adding a straight waveguide in the middle of ring resonators makes coupling length more controllable. Vernier effect in this design could broaden the FSR value while not increasing bending losses. Both through port insertion loss and interstitial peak suppression could be improved by the addition of contra-directional couplers. In our work, several types of racetrack resonators are designed and fabricated to validate the vernier effect and its FSR--extending performance. Parameters (inter-stage coupling, coupling length $\&$ coupling strength of straight waveguide, field transmission, etc.) in design are varied to find out their influence on the performance.
 
\section{Modeling}\label{sec2}
\subsection{Cascaded Coupled Resonator} 

\begin{figure}
\centering
\includegraphics[scale=0.4]{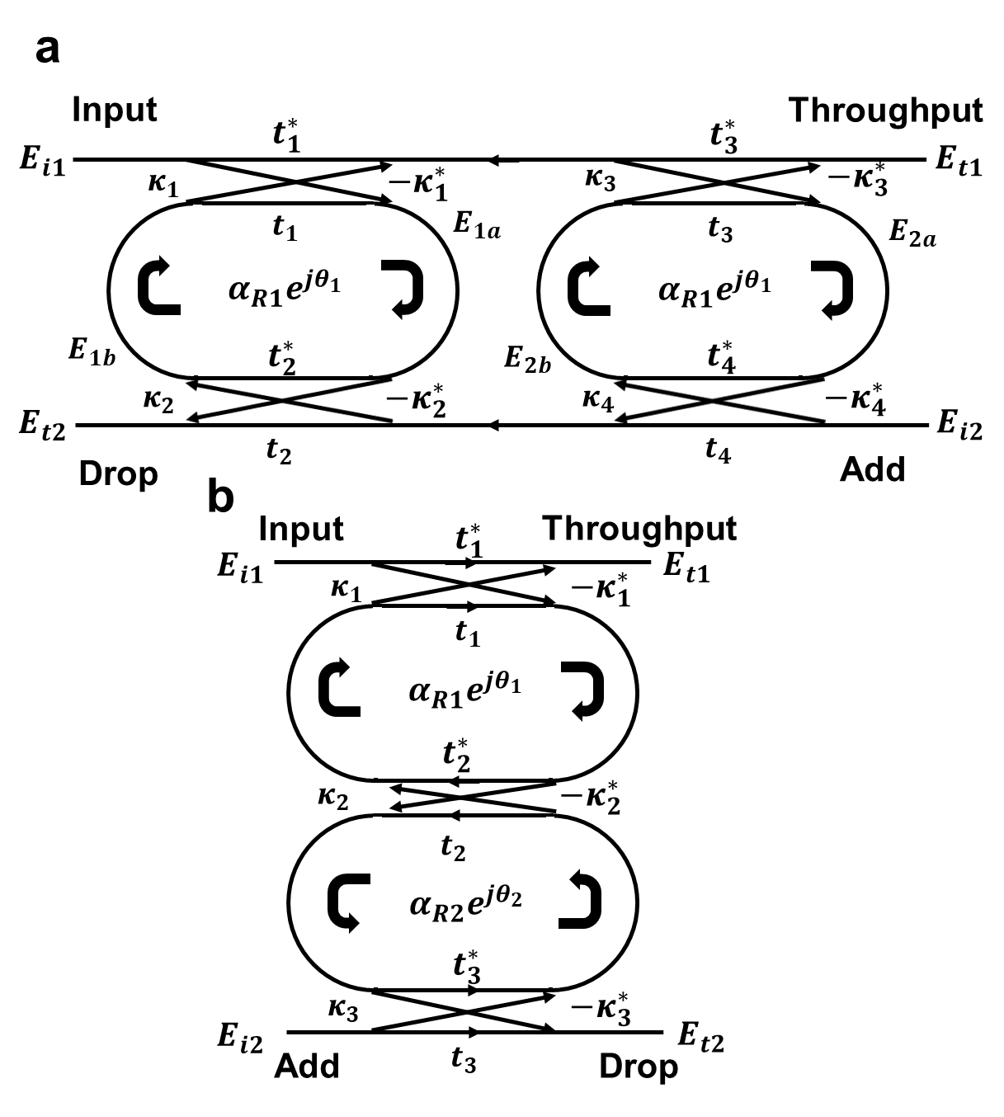}
\caption{Coupling scheme: (a) series pattern. (b) parallel pattern.}
\label{fig:RingScheme}
\end{figure}

To suppress bending loss, radius of racetrack resonators is always settled as large value, which contributes to small FSR. To extend FSR, multiple racetrack resonators are connected to exhibit vernier effect. Two cascaded racetrack resonators are displayed in Fig.\ref{fig:RingScheme} a), which show vernier effect.

The drop port transmission function \cite{rabus2007integrated} in amplitude forms for through port and drop port is expressed by Eq. \ref{con:E1}:

\begin{equation}
(\frac{E_{t1}}{E_{i1}})_{through} =
	\frac{-t_{1}\kappa_{1}^2\alpha_{1}e^{j\theta_{1}}(t_{3}\alpha_{2}e^{j\theta_{2}}-t_{2})}
	{ 1-t_{3}t_{2}\alpha_{2}e^{j\theta_{2}}-t_{2}t_{1}\alpha_{1}e^{j\theta_{1}}+t_{3}t_{1}\alpha_{1}e^{j\theta_{1}}e^{j\theta_{2}} 	}. \label{con:E1}
\end{equation}

\begin{equation}
(\frac{E_{t1}}{E_{i1}})_{drop} =
	\frac{\kappa_{3}\kappa_{2}\kappa_{1}\alpha_{1}\alpha_{2}e^{j\frac{\theta_{1}}{2}}e^{j\frac{\theta_{2}}{2}}}   
	{ 1-t_{3}t_{2}\alpha_{2}e^{j\theta_{2}}-t_{2}t_{1}\alpha_{1}e^{j\theta_{1}}+t_{3}t_{1}\alpha_{1}e^{j\theta_{1}}e^{j\theta_{2}} 	}.
\end{equation}

Where $\kappa_{1}$, $\kappa_{2}$, $\kappa_{3}$ and its conjugated complex form value represents corresponding coupling coefficient, and t$_{1}$, t$_{2}$, t$_{3}$, and its conjugated complex form represents corresponding field transmission factor. $\alpha$ (zero loss: $\alpha$=1) is the total field loss coefficient. To simplify the model, we let $\kappa_{1}$=$\kappa_{3}$.

Series cascaded rings/racetrack shapes with different radii makes it possible to extend FSR to the least common multiple of single one. This is because the resonant condition is satisfied only when both single resonators are in resonance. This leads to the suppression of some transmission peaks of racetrack resonators. The FSR of cascaded resonators [6] with different radii is given by Eq. \ref{con:FSR1} - Eq. \ref{con:FSR2}:

\begin{equation}
FSR = N\times FSR_{1}=M\times FSR_{2} \label{con:FSR1}
\end{equation}  

\begin{equation}
FSR =  |M-N|  \frac{FSR_{1}\times FSR_{2}}{FSR_{1} - FSR_{2}} \label{con:FSR2}
\end{equation}   

Where N and M are natural and coprime numbers. 
The group index n$_{g}$ and effective index n$_{eff}$ is defined as:

\begin{equation}
n_{g} =  n_{eff}-\lambda \frac{\partial n_{eff}}{\partial \lambda}
\end{equation}

\subsection{Serially Coupled Double Resonator} 
To realize box-like filter response, parallelly coupled racetrack resonators are designed and analyzed in Suzuki et al. (1995). The parallel type of racetrack resonator is displayed in Fig. \ref{fig:RingScheme} b). 


This kind of configuration can be treated as gratings. Choosing $\Lambda$ to be odd multiple of a quarter wavelength, constructive interference forms by reflected light. Let $\lambda_0$ to be the center wavelength (1550 $nm$), $\Lambda$ is expressed as Eq. \ref{con: lambda1}: 

\begin{equation}
\Lambda =  (2m+1)\frac{\lambda_{0}}{4n_{eff}}  \label{con: lambda1}
\end{equation}

The parallel coupled racetrack resonators exhibit vernier effect when the periods of FSR of racetrack resonators and gratings follow Eq. \ref{con: lambda2} and Eq. \ref{con: lambda3}. In this case, transmission peaks are suppressed as well, overall FSR is extended.

\begin{equation}
\Lambda = N_{Racetrack}FSR_{Racetrack}=M_{Grating}FSR_{Grating}  \label{con: lambda2}
\end{equation}

\begin{equation}
\Lambda =  \frac{M_{grating}n_{Weff}}{N_{Racetrack}n_{Weff}}\pi r   \label{con: lambda3}
\end{equation}

\begin{figure}
\centering
\includegraphics[scale=0.5]{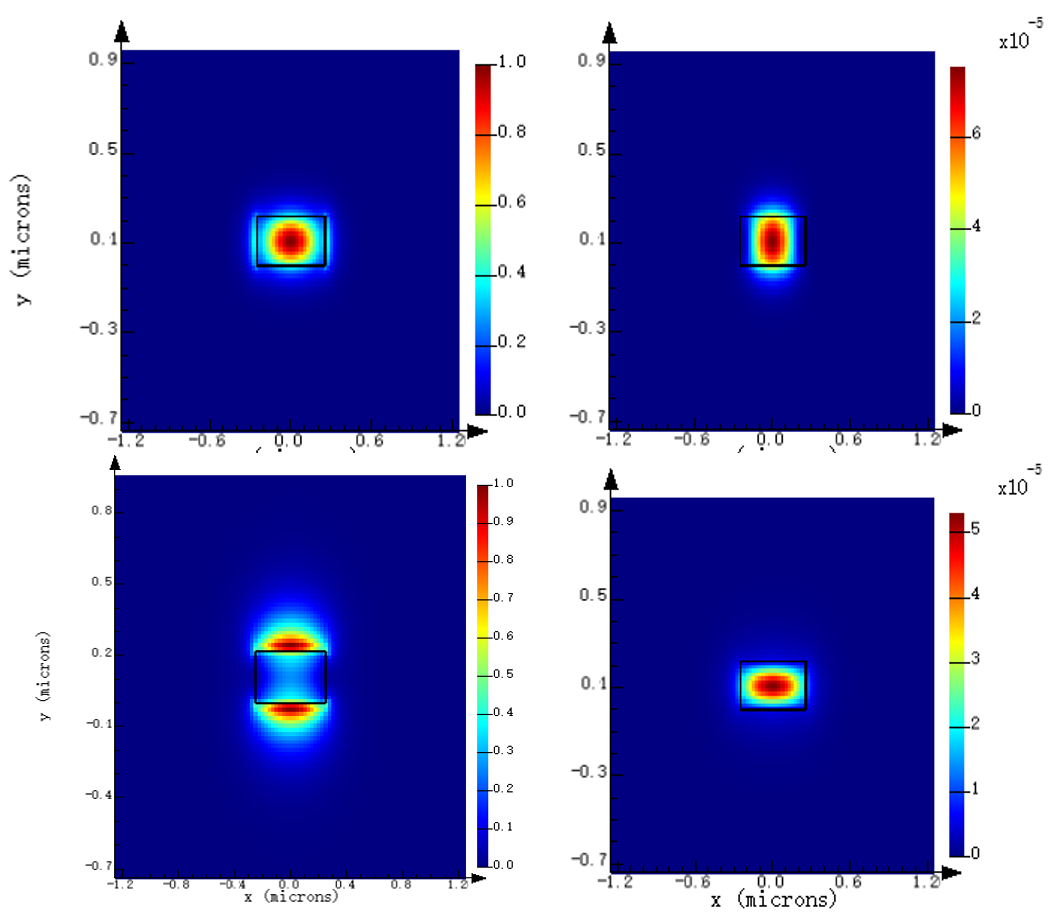}
\caption{Mode files for fundamental TE and TM mode. (a) TE mode: E intensity. (b) TE mode: H intensity. (c) TM mode: E intensity. (d) TM mode: H intensity.}
\label{fig:MODE}
\end{figure}

\section{Results and Discussion}

\subsection{Simulation results}

To determine the effective size of the cascaded racetrack resonator, a strip waveguide with 550~nm width and 220~nm height is modeled to explore the field distribution. The effective index ($\lambda = 1550 nm$) calculated by Taylor expansion with the simulation results have a good agreement~\cite{chrostowski2015silicon}. 

To further decide the parameter of the racetrack resonator, the relationship between inter-coupling length, gap, and coupling coefficient should be explored. We use FDTD simulation to figure out how coupling coefficients depend on these parameters. 
To satisfy the requirement of adequate interstitial suppression, $M$ or $N$ should be small enough. The length of straight waveguide parts in cascaded ring resonators determines the coupling coefficients and eventual FSR. With the unavoidable fabrication error of the lithography process, more copies with slight variations around the appointed circuit size are submitted to analyze and predict the effect of the error (Fig. \ref{fig:Comparision_RingNumber_radisu}a). 

\subsection{Experiment results}
\begin{figure}
\centering
\includegraphics[scale=0.18]{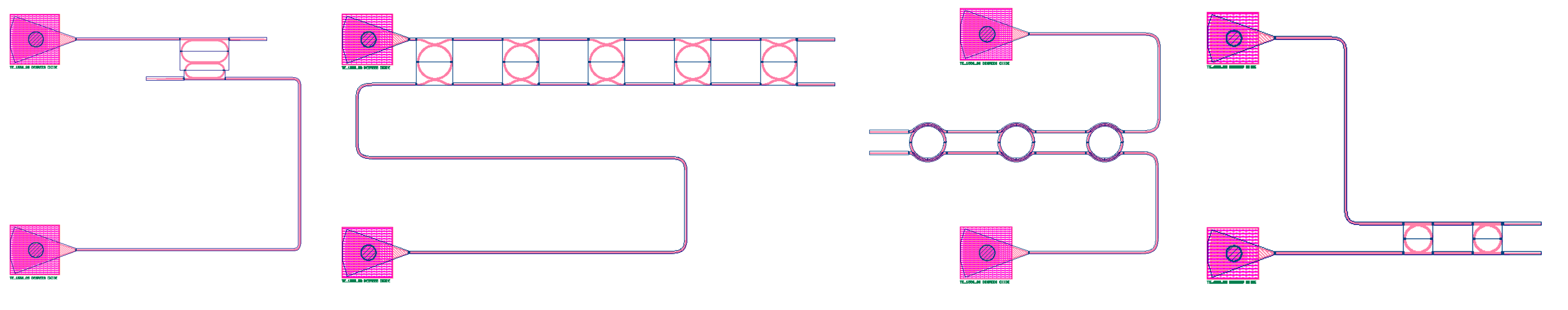}
\caption{Coupling types and devices. (a) Racetrack resonator with a straight coupling.  (b) Parallel ring resonators (5 poles) with arc coupling. (c) Parallel ring resonators (3 poles) with bend coupling. (d) Parallel ring resonators (2 poles) with a straight coupling.}
\label{fig:coupling}
\end{figure}

\begin{figure}
\centering
\includegraphics[scale=0.3]{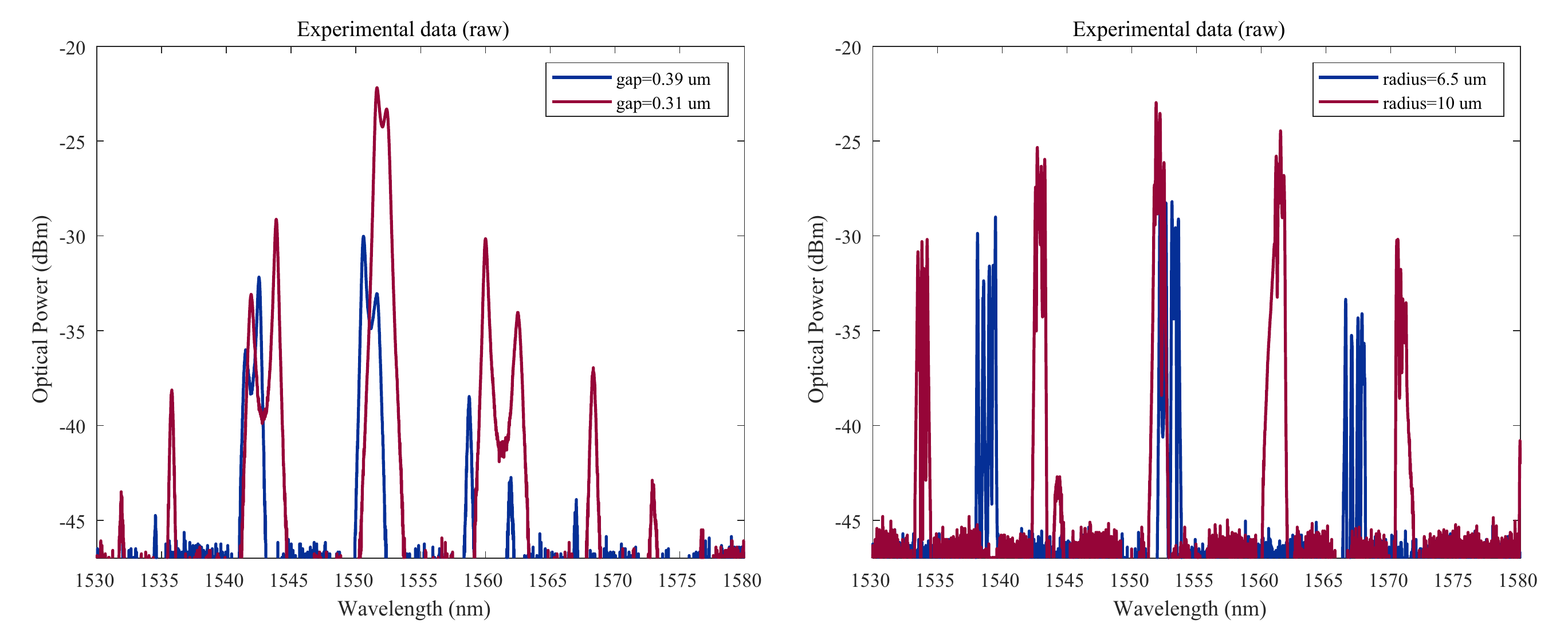}
\caption{Comparison between devices: (a) Experimental results for racetrack resonators. Gaps between resonators are $0.31 \mu m$ (red line) and $0.39 \mu m$ (blue line). (b) Comparison between parallel rings (5 poles) with different radius: $10 \mu m$ (red line) and $6.5 \mu m$ (blue line). }
\label{fig:GAP_5radius_copy_bendstraight}
\end{figure}

\begin{figure}
\centering
\includegraphics[scale=0.3]{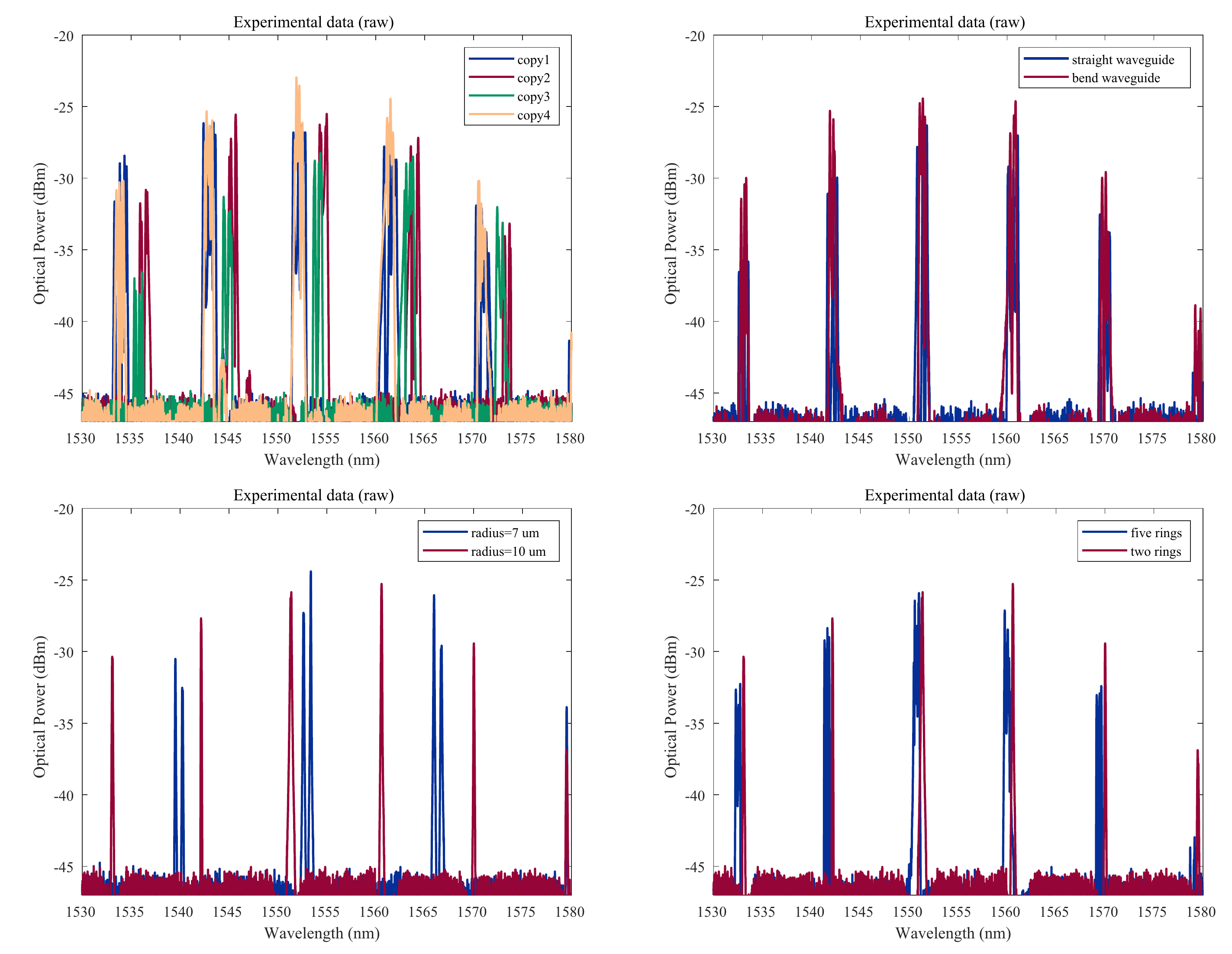}
\caption{Comparison between parallel ring performance. (a) Parallel rings (5 poles) have the same size but display different performances. (b) Parallel rings (3 poles) with different coupling types: straight waveguide (blue) and bend waveguide (red). (c) Parallel rings (2 poles) with different radius : radius=$7 \mu m$ (blue line) and radius=$10 \mu m$ (red line).  (d) Different ring numbers loaded between sources: five rings (blue) and two rings (red).}
\label{fig:Comparision_RingNumber_radisu}
\end{figure}

To limit insertion loss and ensure TE-polarization with a single mode, periodic grating couplers with the fixed-length taper are utilized to couple light from the fiber array to the waveguide. Different coupling types between the waveguide and ring are investigated. Various coupling length has also been tested, given the machining accuracy and fabrication error. Finally, the performance of several parallel rings and racetrack resonators has been verified. For racetrack resonators, decreasing the coupling gap will increase the coupling coefficient and decrease insertion loss. This will also lead to the appearance of parasitic bands and decay of interstitial resonance suppression as in Fig. \ref{fig:GAP_5radius_copy_bendstraight}a. For the parallel ring resonators, we have fabricated three types of coupling: textbf{arc waveguide}, textbf{bend waveguide}, and textbf{straight waveguide}, displayed in Fig. \ref{fig:coupling}. Straight coupling is the most normal way of coupling, while bend coupling can provide stronger coupling at the same condition. Arc coupling provides the weakest strength. Radius in the parallel rings decides the FSR, as Fig. \ref{fig:GAP_5radius_copy_bendstraight}b shows, a larger radius presents a shorter FSR. The insertion loss is smaller as well. To investigate the influence of fabrication error on the final performance, we fabricated and measured four copies. In parallel rings, the coupling type doesn’t affect performance too much, as shown in Fig. \ref{fig:Comparision_RingNumber_radisu}b. For two rings, a larger radius brings in smaller FSR, as shown in Fig. \ref{fig:Comparision_RingNumber_radisu}c, where the ring with $10 \mu m$ generates $10nm$ FSR, while the ring with $7 \mu m$ generates $13nm$ FSR. When more rings are loaded between ports, the center frequency moves toward to lower band, as shown in Fig. \ref{fig:Comparision_RingNumber_radisu} d).

\section{Conclusion}
Our work focused on the vernier effect of cascaded and series-coupled MRRs sensors with different parameters. From our simulation and experiment results, it is proven to be effective for low intra-channel crosstalk at higher data rates by offering larger input-to-through suppression for future applications.

\section*{Acknowledgment}

We acknowledge the Natural Sciences and Engineering Research Council of Canada (NSERC) Silicon Electronic-Photonic Integrated Circuits (SiEPIC) Program, the NSERC CREATE in Quantum Computing program, and the Canadian Microelectronics Corporation (CMC). Devices were fabricated at Advanced Micro Foundry (AMF) A STAR foundry in Singapore. 

\vspace{12pt}
\bibliographystyle{IEEEtran}
\bibliography{main}
\end{document}